# Exploring the atmosphere using smartphones

## Martín Monteiro[a], Patrik Vogt[b], Cecilia Stari[c], Cecilia Cabeza[d], Arturo C. Marti[e],

[a] Universidad ORT Uruguay; monteiro@ort.edu.uy
[b] University of Education Freiburg, Germany; patrik.vogt@ph-freiburg.de
[c] Universidad de la República, Uruguay, cstari@fing.edu.uy
[d] Universidad de la República, Uruguay, cecilia@fisica.edu.uy
[e] Universidad de la República, Uruguay, marti@fisica.edu.uy

The characteristics of the inner layer of the atmosphere, the troposphere, are determinant for the earth's life. In this experience we explore the first hundreds of meters using a smartphone mounted on a quadcopter. Both the altitude and the pressure are obtained using the smartphone's sensors. We complement these measures with data collected from the flight information system of an aircraft. The experimental results are compared with the International Standard Atmosphere and other simple approximations: isothermal and constant density atmospheres.

**The international standard atmosphere**

The atmospheric conditions exhibit strong variations at different points and different times. To provide a unified frame of reference, an atmospheric model, the International Standard Atmosphere (ISA), has been established[1]. It consists of tables of pressure, temperature, density and other variables suitable at mid latitudes over a wide range of altitudes. The ISA is used for several purposes ranging from altimeter calibration to comparison of aircraft performance among others. The ISA is divided into layers with simple temperature variations. The inner layer is the troposphere, from the surface until 11 km of height, in which the temperature presents a linear gradient, named lapse rate, C= - 0.0065 K/m. The pressure $p$ as a function of the height $h$ according to the ISA is shown in Fig. 1.

We also consider two additional atmospheric models. Firstly, a rather crude approximation consists of considering the temperature constant in the inner layers. This model is called isothermal atmosphere. The flight information system usually available to the passengers in many aircrafts provides an clever way to quantify the relationship between temperature and pressure. In Fig. 2 we plot the temperature as a function of the altitude using data collected by a passenger. The linear fit, and g gives a temperature gradient of about C= - 0.0072 K/m. Of course, this value highly depends on the particular atmospheric conditions during this flight and does not necessarily represents accurately the ISA. This exercise can be proposed as a homework to students about to travel on a plane.

The second approximation is obtained neglecting air density variations. Under this approximation valid at small altitudes (a few kilometers), ρ is constant and the atmospheric pressure obeys the hydrostatic equation

$$p = p_0 - \rho g h,$$

where $p_0$ is the pressure at $z=0$. For the sake of comparison, in Fig. 1, the isothermal atmosphere and the constant density model are shown.

**The experiment**

In the present experiment we use the pressure sensor and the GPS of a smartphone attached to a quadcopter. Other interesting experiments involving pressure sensors or GPS can be found in Refs. [2] and [3], the use of quadcopters in teaching physics has been recently considered in [4].

A smartphone, *LG* model *G3* was mounted in a *DJI Phantom 2* using an armband case as seen in Figure 3. The quadcopter was raised until 250 *m* with respect to the take off point, kept hovering a dozen of second and taken down. During the flight, the Androsensor app was used to register the atmospheric pressure using the built-in barometer and the height obtained from the GPS. As the response time of the GPS is rather slow, during ascend and descend the operator tried to control the thrust of the device in such a way that the vertical speed was roughly constant. According to the information provided by the manufacturer of the quadcopter, the maximum ascent speed is 6 m/s and the maximum descent speed is 2 m/s. Care should be taken to avoid damages in persons, animals or properties and also to fulfill the air traffic regulations[5].

**Results**

In figure 4, the pressure as a function of the altitude is shown. From the slope of the linear fit the air density is found to be $\rho = 1.12$ kg/m$^3$. This value is a good approximation of the standard value valid in the inner atmospheric layer.

**Conclusion**

The use of a quadcopter and a smartphone allowed to record the main atmospheric variables. In this experiment, we obtained pressure profiles and, through linear fits, values of the density. This experiment helps to get an insight into the characteristics of the atmosphere using accessible tools.

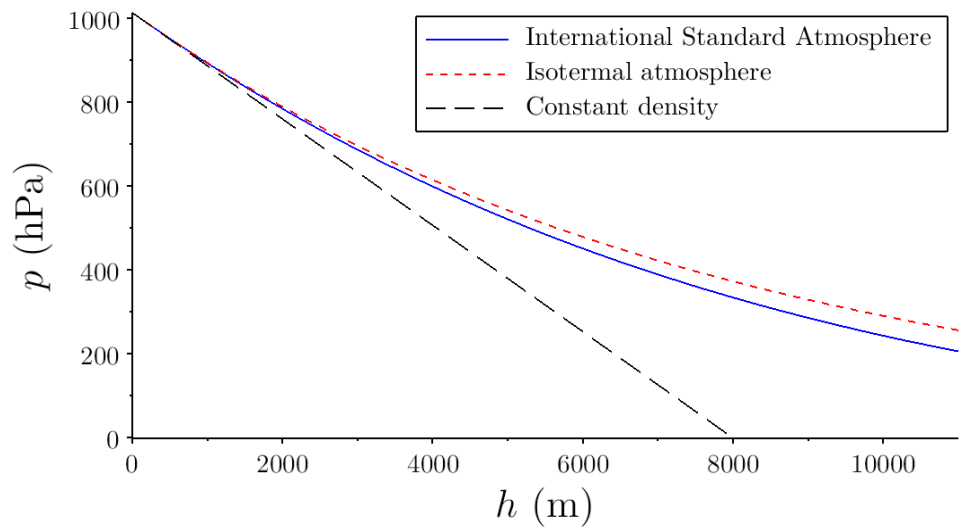

Figure 1. The international standard atmosphere.

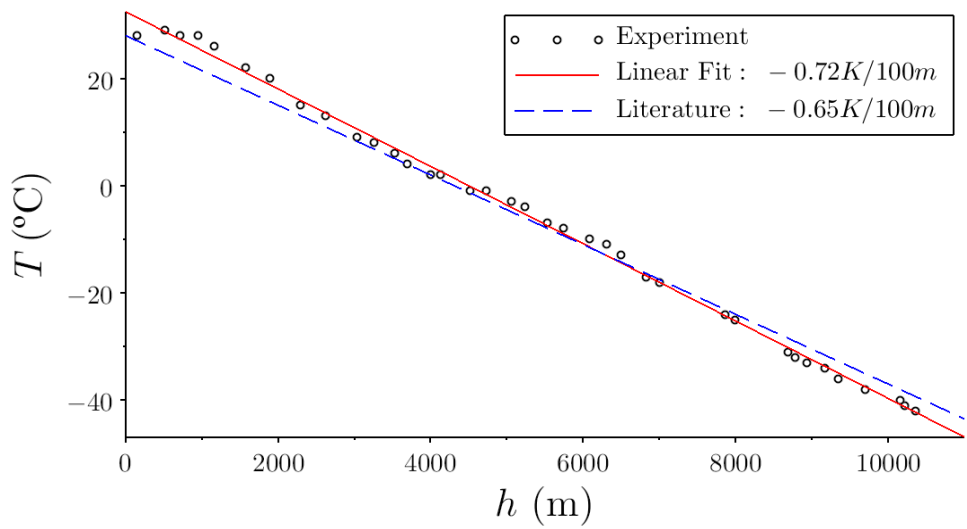

Figure 2. Temperature as a function of the altitude using an aircraft's information system.

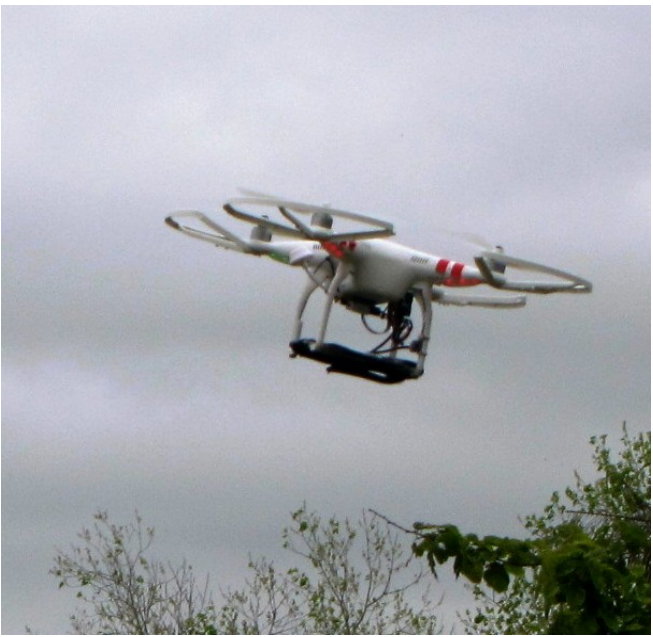

Figure 3. Smartphone mounted on a DJI Phantom 2 using an armband.

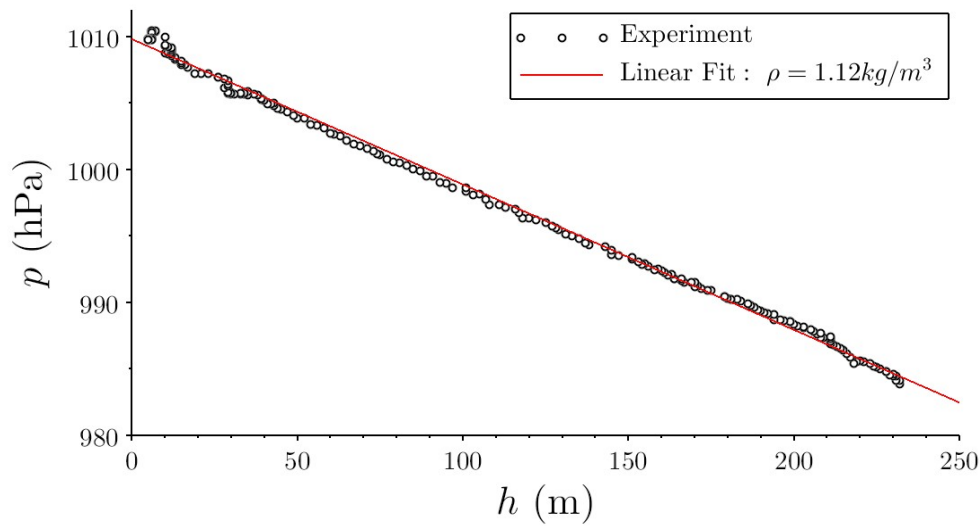

Figure 4. Atmospheric pressure as a function of the altitude recorded by the smartphone.